# HOW TO TWIRL A HULA-HOOP

**A.P. Seyranian\*, A.O. Belyakov†**
Institute of Mechanics
Moscow State Lomonosov University
Michurynski pr. 1, Moscow 119192, Russia
\*seyran@imec.msu.ru  †a_belyakov@inbox.ru

ABSTRACT. We consider twirling of a hula-hoop when the waist of a sportsman moves along an elliptic trajectory close to a circle. For the case of the circular trajectory, two families of exact solutions are obtained. Both of them correspond to twirling of the hula-hoop with a constant angular speed equal to the speed of the excitation. We show that one family of solutions is stable, while the other one is unstable. These exact solutions allow us to obtain approximate solutions for the case of an elliptic trajectory of the waist. We demonstrate that in order to twirl a hula-hoop one needs to rotate the waist with a phase difference lying between $\pi/2$ and $\pi$. An interesting effect of inverse twirling is described when the waist moves in opposite direction to the hula-hoop rotation. The approximate analytical solutions are compared with the results of numerical simulation.

I. INTRODUCTION. A hula-hoop is a popular toy – a thin hoop that is twirled around the waist, limbs or neck. In the last decades it is widely used as an implement for fitness and gymnastic performances, Fig. 1. To twirl a hula-hoop the waist of a sportsman carries out a periodic motion in the horizontal plane. For the sake of simplicity we assume that the waist is a circle and its center moves along an elliptic trajectory close to a circle. Earlier the simple case when a hula-hoop is treated as a pendulum with the pivot oscillating along a line has been considered.[1,2] The stationary rotations of a hula-hoop excited in two directions have been studied by an approximate method of separate motions.[3] Similar problem on the spinner mounted loosely on a pivot with a prescribed bi-directional motion has been treated numerically and experimentally.[4]

    Here we derive the exact solutions in the case of a circular trajectory of the waist center and approximate solutions in the case of an elliptic trajectory. We also check the condition of keeping contact with the waist during twirling. The paper differs from our previous one [5] by new

approximate solutions found analytically and compared with the numerical simulation results.

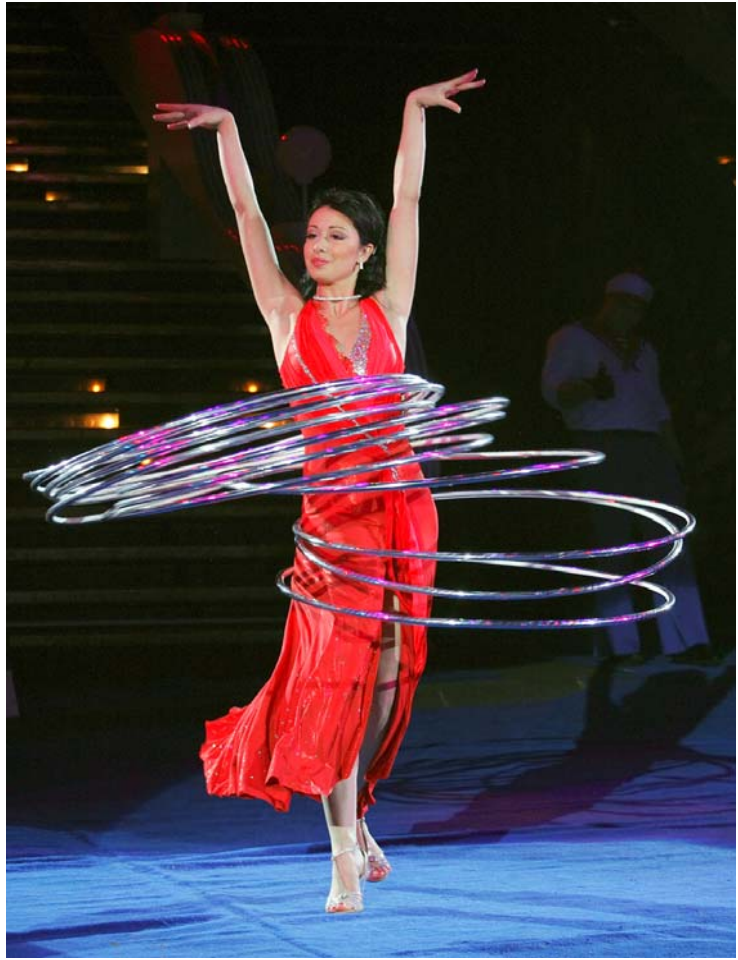

Fig. **1**. Twirling of hula-hoops in the circus performance, Moscow 2008
(photo by K. Stukalov).

II. MAIN RELATIONS AND EXACT SOLUTIONS. We assume that the center $O'$ of a sportsman's waist moves in time $t$ according to the elliptic law $x = a\sin\omega t$, $y = b\cos\omega t$ with the amplitudes $a$, $b$ and the excitation frequency $\omega > 0$, Fig 2. The equations of motion in the waist-fixed coordinate system with allowance for linear viscous damping take the form

$$I_c\ddot{\theta} + k\dot{\theta} = -F_T R \qquad (1)$$

$$m(R-r)\ddot{\varphi} = m(\ddot{x}\sin\varphi + \ddot{y}\cos\varphi) + F_T \qquad (2)$$

$$m(R-r)\dot{\varphi}^2 = N + m(\ddot{x}\cos\varphi - \ddot{y}\sin\varphi) \qquad (3)$$

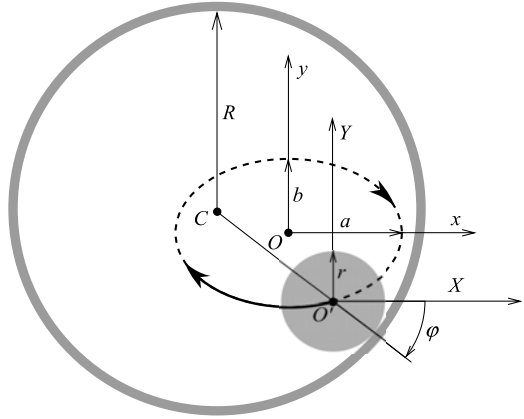

Fig.2. A hula-hoop with the radius $R$ twirling with the angle $\varphi$ around a circular waist (shaded) with the radius $r$. The center $O'$ of the waist moves along the elliptic curve $x = a\sin\omega t$, $y = b\cos\omega t$ with the fixed center $O$.

where $\theta$ is the rotation angle around center of mass $C$, $I_c = mR^2$ is the central moment of inertia of the hula-hoop, $\varphi$ is the angle between axis $x$ and radius $CO'$, $F_T$ is the friction force, $m$ and $R$ are the mass and radius of the hula-hoop, $k$ is the coefficient of viscous friction, $r$ is the radius of the waist, and $N$ is the normal force. Equation (1) describes rotation of the hula-hoop around $C$, while (2) and (3) are equations of motion of the hula-hoop in the longitudinal and transverse directions to the radius $CO'$. Assuming that slipping at the point of contact is absent we obtain the kinematic relation

$$(R-r)\dot{\varphi} = R\dot{\theta} \qquad (4)$$

We exclude from equations (1) и (2) the force $F_T$ and with relation (4) obtain the equation

$$\ddot{\varphi} + \frac{k}{2mR^2}\dot{\varphi} + \frac{\omega^2(a\sin\omega t\sin\varphi + b\cos\omega t\cos\varphi)}{2(R-r)} = 0 \qquad (5)$$

From equation (3) we find the normal force and imply the condition $N > 0$ as
$$m(R-r)\dot{\varphi}^2 + m\omega^2(a\sin\omega t \cos\varphi - b\cos\omega t \sin\varphi) > 0 \qquad (6)$$
which means that the hula-hoop during its motion keeps contact with the waist of the sportsman.

We introduce new time $\tau = \omega t$ and non-dimensional parameters
$$\gamma = \frac{k}{2mR^2\omega}, \quad \alpha = \frac{a}{2(R-r)}, \quad \beta = \frac{b}{2(R-r)} \qquad (7)$$
Then equation (5) and inequality (6) take the form
$$\ddot{\varphi} + \gamma\dot{\varphi} + \alpha\sin\tau\sin\varphi + \beta\cos\tau\cos\varphi = 0 \qquad (8)$$
$$\dot{\varphi}^2 + 2(\alpha\sin\tau\cos\varphi - \beta\cos\tau\sin\varphi) > 0 \qquad (9)$$
where the dot means differentiation with respect to the time $\tau$. These equations contain three non-dimensional parameters: the damping coefficient $\gamma$, and the amplitudes $\alpha$ and $\beta$ along the axes *x* and *y*. Relation between the amplitudes $\alpha$ and $\beta$ determines the form of ellipse – the trajectory of the waist center. For $\beta = 0$ the trajectory is a line, and for $\alpha = \beta$ it is a circle.

For the equal excitation amplitudes $\alpha = \beta$ and arbitrary damping coefficient $\gamma$ equation (8) takes the form
$$\ddot{\varphi} + \gamma\dot{\varphi} + \alpha\cos(\varphi - \tau) = 0 \qquad (10)$$
which has the exact solution [5]
$$\varphi = \tau + \psi \qquad (11)$$
with the constant initial phase $\psi$ given by the equation
$$\cos\psi = -\gamma/\alpha \qquad (12)$$
Therefore, solution (11) exists only under the condition
$$|\gamma| \le |\alpha| \qquad (13)$$
From equation (12) we find
$$\psi = \pm\arccos(-\gamma/\alpha) + 2\pi k, \quad k = 0,1,2,... \qquad (14)$$
Solutions (11), (14) correspond to the rotation of the hula-hoop with the constant angular velocity equal to the excitation frequency $\omega$.

Let us investigate the stability of the obtained solutions. For this purpose we take the angle $\varphi$ in the form $\varphi = \tau + \psi + \eta(\tau)$ where $\eta(\tau)$ is a small quantity, and substitute it into equation (10). Taking linearization with respect to $\eta$ and with the use of (12) we obtain a linear equation

$$\ddot{\eta} + \gamma \dot{\eta} - \alpha \sin \psi \, \eta = 0 \qquad (15)$$

Without loss of generality we assume $\alpha > 0$ since the case $\alpha < 0$ is reduced to the previous one by the time transformation $\tau' = \tau + \pi$ in equation (8). According to Lyapunov's theorem on the stability based on a linear approximation [6] solution (11), (14) is asymptotically stable if all the eigenvalues of linearized equation (15) have negative real parts. From Routh-Hurwitz criterion [6] due to the assumption $\alpha > 0$ we obtain the stability conditions as

$$\gamma > 0, \quad \sin \psi < 0 \qquad (16)$$

From conditions (16) and relation (14) we find that for $0 < \gamma < \alpha$ solution (11) with

$$\psi = -\arccos(-\gamma/\alpha) + 2\pi k, \; k = 0,1,2,... \qquad (17)$$

is asymptotically stable, and solution (11) with

$$\psi = \arccos(-\gamma/\alpha) + 2\pi k, \; k = 0,1,2,... \qquad (18)$$

is unstable. For negative damping $\gamma < 0$ both solutions (11), (17) and (11), (18) are unstable.

It is necessary to verify for the exact solutions (11), (14) the condition of twirling without losing contact (9) which takes the form

$$1 - 2\alpha \sin \psi > 0 \qquad (19)$$

Due to (16) and the assumption $\alpha > 0$ condition (19) is satisfied for the stable solution (11), (17).

Thus, solution (11), (17) under the condition $0 < \gamma < \alpha$ provides asymptotically stable twirling of the hula-hoop with the constant angular velocity $\omega$ without losing contact with the waist of the sportsman. The phase of this solution belongs to the interval $-\pi \leq \psi \leq -\pi/2$ plus $2\pi k$ to both sides of this inequality, and for vanishing damping $\gamma \to +0$ the phase tends to $-\pi/2$. Below we will show that this phase inequality also holds for the approximate solutions. This is the way how to twirl a hula-hoop!

III. APPROXIMATE SOLUTIONS. For close but not equal amplitudes $\alpha \approx \beta$ we introduce notation $\varepsilon = (\alpha - \beta)/2$, $\mu = (\alpha + \beta)/2$. For the sake of simplicity we assume $\alpha \geq |\beta|$ which means, $\varepsilon \geq 0$ $\mu \geq 0$. Then equations (8), (9) take the form

$$\ddot{\varphi} + \gamma\dot{\varphi} + \mu\cos(\varphi - \tau) = \varepsilon\cos(\varphi + \tau) \qquad (20)$$

$$\dot{\varphi}^2 - 2\mu\sin(\varphi - \tau) + 2\varepsilon\sin(\varphi + \tau) > 0 \qquad (21)$$

Taking $\varepsilon$ as a small parameter we apply perturbation method assuming that solution of (20) can be expressed in a series

$$\varphi_s(\tau) = \varphi_*(\tau) + o(\varepsilon), \quad \varphi_*(\tau) = \tau + \varphi_0 + \varepsilon\varphi_1(\tau), \qquad (22)$$

where $\varphi_*(\tau)$ is the first approximation function. After substitution of series (22) in (20) and grouping the terms by equal powers of $\varepsilon$ we derive the following chain of equations

$$\begin{aligned}\varepsilon^0: &\quad \gamma + \mu\cos\varphi_0 = 0 \\ \varepsilon^1: &\quad \ddot{\varphi}_1 + \gamma\dot{\varphi}_1 - \mu\sin(\varphi_0)\varphi_1 = \cos(\varphi_0 + 2\tau)\end{aligned} \qquad (23)$$

Taking solution of the first equation (23)

$$\varphi_0 = -\arccos(-\gamma/\mu) + 2\pi k \qquad (24)$$

corresponding to the stable zero order approximation $\varphi(\tau) = \tau + \varphi_0$, we write the second equation (23) as

$$\ddot{\varphi}_1 + \gamma\dot{\varphi}_1 + \sqrt{\mu^2 - \gamma^2}\,\varphi_1 = \cos(\varphi_0 + 2\tau) \qquad (25)$$

It has a unique periodic solution

$$\varphi_1(\tau) = C\sin(2\tau + \varphi_0) + D\cos(2\tau + \varphi_0) \qquad (26)$$

where

$$C = \frac{2\gamma}{\mu^2 + 3\gamma^2 - 8\sqrt{\mu^2 - \gamma^2} + 16}, \quad D = \frac{-4 + \sqrt{\mu^2 - \gamma^2}}{\mu^2 + 3\gamma^2 - 8\sqrt{\mu^2 - \gamma^2} + 16} \qquad (27)$$

We see that the approximate solutions $\varphi_*(\tau)$ with (24), (26), (27) differ from the exact solutions (11), (17) by small vibrating terms and phase shift. Note that the approximate solutions were obtained with the assumption that the excitation amplitudes and damping are not small.

To find the stability conditions for solution (22), (24), (26) we take a small variation to the solution of (22) $\varphi = \varphi_s + u$ and substitute this expression into (20). After linearization with respect to $u$ and keeping only terms of first order we obtain a linear equation

$$\ddot{u} + \gamma\dot{u} + [-\mu(\sin\varphi_0 + \varepsilon\varphi_1\cos\varphi_0) + \varepsilon\sin(2\tau + \varphi_0)]u = 0 \qquad (28)$$

where $\varphi_0$ is given by expression (24). Equation (28) can be written in the form of damped Mathieu-Hill equation as

$$\ddot{u} + \gamma \dot{u} + [p + \varepsilon \Phi(2\tau)]u = 0 \qquad (29)$$

where

$$p = \sqrt{\mu^2 - \gamma^2}, \quad \Phi = (\gamma C + 1)\sin(2\tau + \varphi_0) + \gamma D \cos(2\tau + \varphi_0) \qquad (30)$$

Then the stability condition (absence of parametric resonance at all frequencies $\sqrt{p}$) is given by the inequalities [6]

$$\varepsilon < \frac{2\gamma}{\sqrt{(\gamma C + 1)^2 + \gamma^2 D^2}} \qquad (31)$$

with $C$ and $D$ defined in (27). This is the inequality to the problem parameters $\gamma$, $\varepsilon$ and $\mu$.

The condition of twirling without losing contact (21) takes the following form

$$\varepsilon < \frac{1 + 2\sqrt{\mu^2 - \gamma^2}}{2} \sqrt{\frac{\mu^2 + 3\gamma^2 - 8\sqrt{\mu^2 - \gamma^2} + 16}{\mu^2 + 8\gamma^2 - 12\sqrt{\mu^2 - \gamma^2} + 36}} \qquad (32)$$

Conditions (31), (32) imply restrictions to $\varepsilon$, i.e. how much the elliptic trajectory of the waist center differs from the circle.

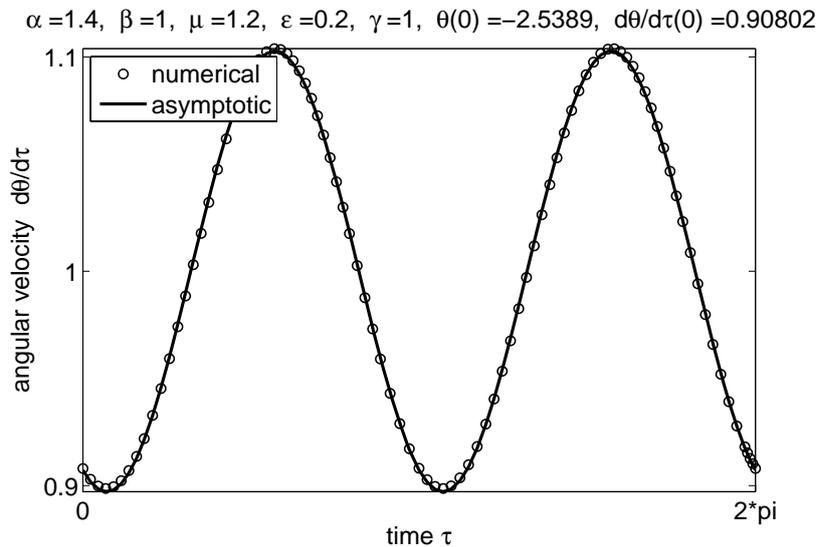

Fig. 3. Comparison between the approximate analytical and numerical results.

In Fig. 3 the approximate analytical solution is presented and compared with the results of numerical simulation for the case when the excitation amplitudes $\alpha, \beta$ and damping coefficient $\gamma$ are not small.

IV. SMALL EXCITATION AMPLITUDES AND DAMPING. It is interesting to consider the case when the excitation amplitudes and damping coefficient are small having the same order as $\varepsilon$. Then we introduce new parameters $\tilde{\mu} = \mu/\varepsilon$ and $\tilde{\gamma} = \gamma/\varepsilon$ and assume that the solution has the form

$$\varphi(\tau) = \rho\tau + \varphi_0(\tau) + \varepsilon\varphi_1(\tau) + o(\varepsilon), \tag{33}$$

where $\rho$ is the angular velocity of rotation, and the functions $\varphi_0(\tau), \varphi_1(\tau)$ are supposed to be limited in time. We substitute expression (33) into (20) and equating terms of the same powers of $\varepsilon$ obtain the following equations

$$\varepsilon^0: \quad \ddot{\varphi}_0 = 0$$
$$\varepsilon^1: \quad \ddot{\varphi}_1 = \cos(\varphi_0 + \tau + \rho\tau) - \tilde{\mu}\cos(\varphi_0 - \tau + \rho\tau) - \tilde{\gamma}\dot{\varphi}_0 - \tilde{\gamma}\rho \tag{34}$$

From the first equation (34) we get $\varphi_0 = const$. Then the second equation (34) can have non-growing solutions only when $\rho$ takes the values -1, 0, 1. Thus, besides clockwise rotation $\rho = 1$ we have also counterclockwise rotation $\rho = -1$, and no rotational solution $\rho = 0$. The letter case is not interesting, so we omit it.

For clockwise rotation $\rho = 1$ from equations (33) and (34) we obtain in the first approximation the solution

$$\varphi(\tau) = \tau + \varphi_0 + \varepsilon\varphi_1(\tau), \tag{35}$$

$$\varphi_1(\tau) = -\frac{1}{4}\cos(\varphi_0 + 2\tau), \quad \cos\varphi_0 = -\frac{\gamma}{\mu} \tag{36}$$

First expression (36) is the special case of (26), (27) for $\mu = \gamma = 0$. To verify the stability conditions for solution (35) we use damped Mathieu-Hill equation (29) and for the case of small damping and excitation amplitudes get $\gamma > 0$, $\sin\varphi_0 < 0$.[6] These conditions are similar with inequalities (16) derived for the undisturbed exact solution. Thus, the stable solution (35) with $\varphi_0 = -\arccos(-\gamma/\mu) + 2\pi k$ exists for $0 < \gamma < \mu$.

For solution (35) condition (21) of keeping contact in the first approximation reads

$$\varepsilon < \frac{1+2\sqrt{\mu^2-\gamma^2}}{3} \qquad (37)$$

For counterclockwise rotation $\rho = -1$ we obtain in the first approximation the solution

$$\varphi(\tau) = -\tau + \varphi_0 + \frac{\mu}{4}\cos(\varphi_0 - 2\tau), \quad \cos\varphi_0 = -\frac{\gamma}{\varepsilon} \qquad (38)$$

with the stability conditions $\gamma > 0$, $\sin\varphi_0 > 0$. Thus, the stable counterclockwise rotation (38) with $\varphi_0 = \arccos(-\gamma/\mu) + 2\pi k$ exists for $0 < \gamma < \varepsilon$. For this case condition (21) takes the form similar with (37)

$$\mu < \frac{1+2\sqrt{\varepsilon^2-\gamma^2}}{3} \qquad (39)$$

Inequalities (37), (39) are satisfied for sufficiently small $\gamma$, $\varepsilon$ and $\mu$.

Stable clockwise and counterclockwise rotations (35), (36) and (38) coexist if $0 < \gamma < \min\{\varepsilon, \mu\}$. Such possibility is demonstrated in the circus performance, see Fig. 1. Coexisting clockwise and counterclockwise rotations are illustrated in Figs. 4 and 5.

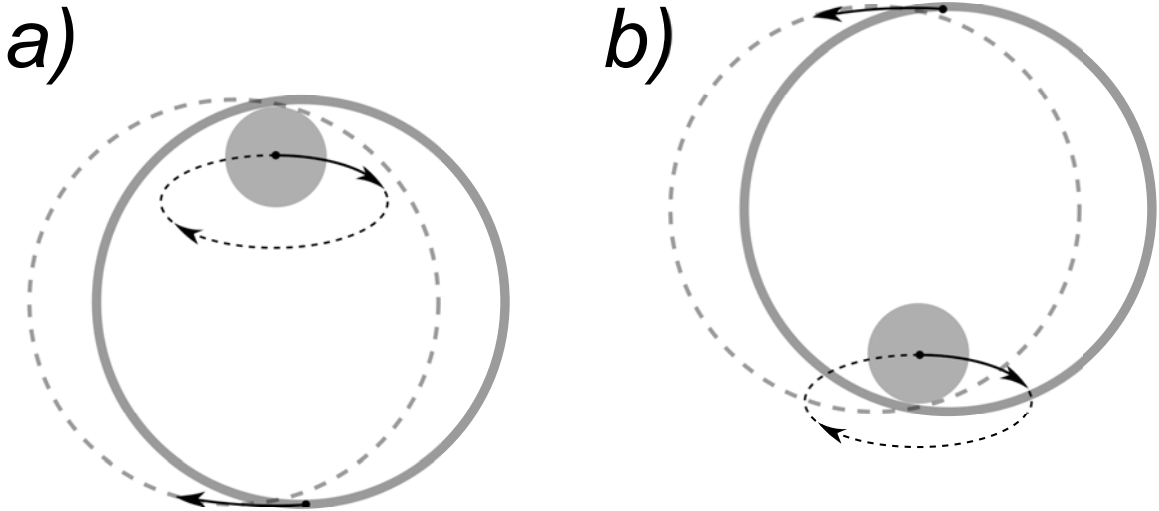

Fig. 4. Stable twirling of the hula-hoop for the cases: a) direct twirling  b) inverse twirling.

Fig. 4 demonstrates phase difference for stable rotations in both directions.

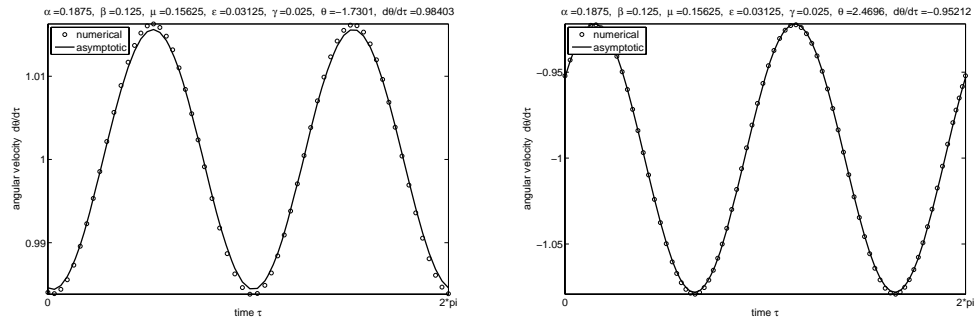

Fig. 5. Comparison between the approximate analytical and numerical results for small excitation amplitudes and damping coefficient for clockwise and counterclockwise rotations.

In Fig. 5 the approximate analytical solutions for rotations in both directions are presented and compared with the results of numerical simulation for the case of small excitation amplitudes $\alpha, \beta$ and the damping coefficient $\gamma$. The values of $\alpha, \beta$ correspond to the dimensional parameters $a = 15$ cm, $b = 10$ cm, $r = 10$ cm, $R = 50$ cm.

V. CONCLUSION. We have derived simple explicit relations showing how to twirl a hula-hoop stably in both directions without losing contact with the waist.